\documentclass[pra,twocolumn,10pt]{revtex4}
\usepackage{amsmath}
\usepackage{dcolumn}
\usepackage[cp1251]{inputenc}
\usepackage[english]{babel}
\usepackage{epsfig}
\usepackage{amssymb}

\newcommand{\sign}{{\rm sign}}

\begin{document}

\title{Berry phase in atom optics}
\author{Polina V. Mironova,$^1$ Maxim A. Efremov,$^{1,2}$  Wolfgang P. Schleich$^1$}
\affiliation{$^1$Institut f\"ur Quantenphysik and Center for
Integrated Quantum Science and Technology ($\it IQ^{ST}$),
Universit\"at Ulm, D-89069 Ulm, Germany \\
$^2$A.M. Prokhorov General Physics Institute, Russian Academy of
Sciences, 119991 Moscow, Russia}

\email{maxim.efremov@uni-ulm.de}

\date{\today}

\begin{abstract}
We consider the scattering of an atom by a sequence of two
near-resonant standing light waves each formed by
two running waves with slightly different wave vectors. 
Due to opposite detunings of the two standing waves and within the
rotating wave approximation, the adiabatic approximation applied to the
atomic center-of-mass motion and a smooth turn-on and -off of the
interaction, the dynamical phase cancels out and the final state of
the atom differs from the initial one only by the sum of the two Berry 
phases accumulated in the two interaction regions.
This phase depends on the position of the atom in a way such that the wave
packet emerging from the scattering region will focus, which
constitutes a novel method to observe the Berry phase without
resorting to interferometric methods.

\end{abstract}
\maketitle

\section{Introduction}

The geometric phase \cite{Berry84,Shapere-Wilczek,Bohm} manifests
itself in many different phenomena of physics ranging from the
polarization change in  the propagation of light in fibers
\cite{Tomita}, via the precession of a neutron in a magnetic field
\cite{Bitter,Wien,Rauch01,Rauch05,Rauch09}, to the quantum dynamics
of dark states  in an atom \cite{Oberthaler}. The geometric phase
has also been used in topological quantum computing \cite{Vedral} as
realized, for example, with trapped ions \cite{Leibfried}. In the
present paper we propose a scheme to observe the geometric phase in
the context of atom optics.

\subsection{Brief review of geometric phases}

The concept of the geometric phase arises in the context of a
Hamiltonian which depends on a parameter which is slowly varying in
time. When this variation is cyclic, that is the Hamiltonian returns
to its initial form, the instantaneous eigenstate will not
necessarily regain its original value, but will pick up a phase.
This phenomenon has been verified in experiments with polarized
light, radio waves, molecules, and many other systems. The most
prominent example is the Aharonov-Bohm effect \cite{Aharonov59},
which was observed in 1959 and interpreted \cite{Berry84} 
in terms of the geometric phase. Moreover, many familiar problems, such as
the Foucault pendulum, or the motion of a charged particle in a
strong magnetic field, usually not associated with the Berry phase,
may be explained elegantly \cite{Shapere-Wilczek,Bohm} in terms of
it.

Since the landmark paper \cite{Berry84} on the geometric phase,
extensive research, both theoretical and experimental, has been
pursued on quantum holonomy \cite{Shapere-Wilczek,Bohm}, adiabatic
\cite{Simon,Samuel,Barut,Hannay,Keck,Tomita,Bitter,Zwanziger} and
non-adiabatic \cite{Moore,Wang}, cyclic \cite{Aharonov,Baudon} and
non-cyclic \cite{Rauch05,Ioffe}, Abelian \cite{Sanders} and
non-Abelian \cite{Sarandy,Zee}, as well as off-diagonal
\cite{Pistolesi,Rauch01} geometric phases. Moreover, geometric phase
effects in the coherent excitation of a two-level atom have been
identified \cite{Barnett,Tewari,Thomaz}. Since geometric phases are
rather insensitive to a particular kind of noise
\cite{Chiara,Rauch09}, they are useful in the construction of robust
quantum gates \cite{Vedral,Uranyan,Matsumoto,Pascazio,Moelmer}.
Although proposals have already been given for the observation of
the geometric phase in atom interferometry \cite{Reich,Oberthaler},
so far only the dependence on the internal atomic degrees of freedom
was investigated. In the present paper we extend this approach by
taking into account external atomic degrees of freedom, that is the
center-of-mass-motion of the atom.

\subsection{Our approach}

We consider the scattering of a two-level atom from a near-resonant
standing light wave formed by two linear polarized running waves
with identical electric field amplitudes and frequencies. The
propagation direction of the two waves is slightly different and the
electromagnetic field is detuned with respect to the resonance
frequency of the atom. Within the Raman-Nath approximation \cite{Sch,Yakovlev} 
on the atomic center-of-mass motion, adiabatic turn-on and -off of the
interaction and with the rotating wave approximation, we obtain a
condition for the cancellation of the dynamical phase and show that
the scattering process is determined solely by the Berry phase
depending on the internal and external atomic degrees of freedom.
The key observation in establishing this condition is the fact that
the dynamical phase is antisymmetric in the detuning, whereas the
geometric phase is symmetric. As a result, a sequence of two such
scattering arrangements which  differ in the sign of their detunings
eliminates the dynamical phase and leads to the sum the two corresponding geometric
phases. To analyze the geometric phase we use the approach \cite{Dalibard} 
based on the adiabatic eigenstates, that is the dressed state picture.

Since the geometric phase imprinted onto the internal state is
position-dependent, we propose a scheme to observe the geometric phase 
based on the narrowing of the atomic wave packet. This
application of the Berry phase might be useful in the realm of 
atom lithography \cite{Oberthaler-Pfau}.

\subsection{Relation to earlier work}

It is for three reasons that our approach is different from earlier
work on the Berry phase arising in the internal dynamics of 
two-level atoms driven by laser fields: (i) in our scheme we
compensate the dynamical phase; (ii) the geometric phase acquired by
the internal states is imprinted onto the center-of-mass motion of the
atoms,  and (iii) our setup does not require a traditional
interference arrangement.

In a landmark experiment the dynamical phase of a neutron precessing
in a magnetic field has been compensated \cite{Rauch09} by an
additional $\pi$-pulse.  In our scheme this cancellation of the
dynamical phase is achieved by changing the sign of the detuning as
the atom interacts with the first and then with the second standing
light field. Moreover, in the experiment described in \cite{Oberthaler} 
the Berry phase is observed in the internal states only and is read out by
interferometry of these states.

We extend these ideas to atom optics
where the center-of-mass motion is treated quantum mechanically.
Here we take advantage of the entanglement between the atomic states
and the center-of-mass motion, which allows us to read out the
information about the geometric phase using the dynamics and the
self-interference of the wave packet.

\subsection{Outline of the article}

Our article is organized as follows. In Sec. II we formulate the
problem addressed in the present paper, define our model and
evaluate the Hamiltonian describing the interaction of a two-level
atom with an appropriately designed standing wave. Next we connect
in Sec. III our model with the one of Ref. \cite{Berry84} and
recall the expressions for the dynamical and geometric phases. Here
we emphasize the connection between the dressed and atomic states.
Since the geometric phase is determined by the path in parameter
space, we construct in Sec. IV the circuit determined by the envelope 
of the electric field. Section V is devoted to
the derivation of explicit expressions for the geometric and
dynamical phases. In particular, we consider the weak field limit
where the Rabi frequency is much smaller than the detuning. 
As an example, we evaluate the geometric and dynamical phases for 
the Eckart envelope in Section VI.
Sections VII and VIII are dedicated to the discussions of the
cancellation of the dynamical phase and the read-out of the geometric
phase with the help of the center-of-mass motion. We summarize our
main results in Section IX.

In order to keep the paper self-contained we have included detailed
calculations in several appendices. For example, 
in Appendix \ref{wkb-method} we apply the WKB-method and 
perturbation theory to rederive the dynamical and geometric phase 
for a two-level atom. Moreover, in Appendix \ref{integrals-details}
we calculate  the integrals determining the geometric and
dynamical phases for a special form of the field envelope, which 
smoothly switches on and off. In this case the path in parameter
space circles many times around the origin. Finally, in Appendix
\ref{flux-evaluation} we evaluate the flux through these infinitely many
windings.

\section{Formulation of the problem}

In the present section we formulate the problem of a two-level atom
scattering off two running electromagnetic waves with almost
opposite wave vectors. For this purpose we first establish the
relevant Hamiltonian and then evaluate the matrix element of the
interaction. We show that under appropriate conditions this quantity
factorizes into a product of three terms, which allows us to imprint
the center-of-mass motion onto a geometric phase.

%%%%%%%%%%%%%%%%%%%%%%%%%%%%%%%%%%%%%%%%%%%%%%%%%%%%%%%%%%%%%%%%%%%%%%%%%%%%%%

\begin{figure}
\includegraphics[width=0.45\textwidth]{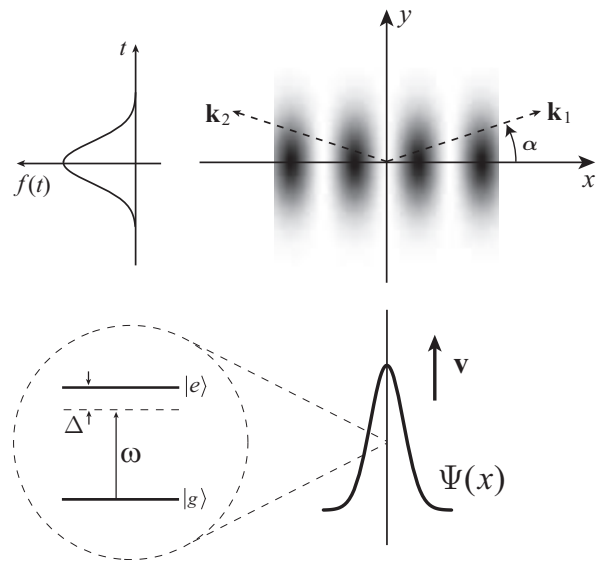}
\caption{\label{fig_setup} Scattering of the wave packet
$\Psi=\Psi(x)$ of a two-level atom by a standing electromagnetic
field formed by two propagating waves of wave vectors $\mathbf{k}_1$
and $\mathbf{k}_2$. The latter assume an angle $\alpha$ and
$\pi-\alpha$ with respect to the $x$-axis. The field envelope along
the $y$-axis translates, according to the relation $y=vt$, into the
time-dependent function $f=f(t)$ as the atom propagates through the
field with velocity $\mathbf{v}=v\,{\rm{\mathbf{e}}}_y$. The
frequency $\omega$ of the field is detuned from the frequency of the
atomic transition between the ground and excited states $|g\rangle$
and $|e\rangle$ by $\Delta$.}
\end{figure}

%%%%%%%%%%%%%%%%%%%%%%%%%%%%%%%%%%%%%%%%%%%%%%%%%%%%%%%%%%%%%%%%%%%%%%%%%%%%%%

\subsection{Set-up}

We consider the scattering of a two-level atom off a near-resonant
standing light field created by  two traveling waves of wave vectors
$\mathbf{k}_1\equiv (k\cos\alpha,k\sin\alpha)$ and
$\mathbf{k}_2\equiv (-k\cos\alpha,k\sin\alpha)$. Both fields form an
angle $\alpha$ relative to the $x$-axis of a Cartesian coordinate
system as shown in Fig. \ref{fig_setup}. The two running waves have
identical wave number, that is $|\mathbf{k}_1|=|\mathbf{k}_2|\equiv
k$, but propagate against each other since
$\left(\mathbf{k}_1\right)_x=-\left(\mathbf{k}_2\right)_x$. As a
result the electric field reads
\begin{equation}
 \label{stwE}
    \mathbf{E}(t,\mathbf{r})=\mathbf{E}_0(\mathbf{r})
    \left[\sin(\mathbf{k}_2\mathbf{r}-\omega t)-\sin(\mathbf{k}_1\mathbf{r}-\omega t)\right],
\end{equation}
where $\mathbf{E}_0=\mathbf{E}_0(\mathbf{r})$ describes the
position-dependent real-valued amplitude of the wave with 
frequency $\omega$.

The time evolution of the state vector $|\Psi\rangle$ of the
two-level atom interacting with the electromagnetic field
$\mathbf{E}$ follows from the Schr\"odinger equation
\cite{Sch,Yakovlev}
\begin{equation}
\label{Sheq}
    i\hbar\frac{d}{d t}|\Psi\rangle=
    \left(\hat{H}_{cm}+\hat{H}_{a}-\hat{\mathbf{d}}\cdot\hat{\mathbf{E}}\right)|\Psi\rangle,
\end{equation}
where the Hamiltonian
\begin{equation}
 \label{Hcm}
    \hat{H}_{cm}\equiv\frac{\hat{\mathbf{p}}^2}{2M}
\end{equation}
is the kinetic energy operator of the center-of-mass motion of the
atom of mass $M$. Here, $\hat{H}_a$ denotes the Hamiltonian of the
free two-level atom with energy eigenstates $|e\rangle $ and
$|g\rangle $ and the corresponding energy eigenvalues
$E_e\equiv\hbar\omega_e$ and $E_g\equiv\hbar\omega_g$, shown in Fig.
\ref{fig_setup}. The field frequency $\omega$ is assumed to be
detuned from the frequency of the atomic transition between
$|g\rangle$ and $|e\rangle$ by
$\Delta\equiv\omega_e-\omega_g-\omega$. The interaction between the
atom and the electromagnetic wave is described by the dipole moment
operator $\hat{\mathbf{d}}$. Moreover, the carrot on $\mathbf{E}$
indicates the operator nature due to the quantum mechanical
description of the motion of the atom.

The mean value of the velocity $v$ of the atom in the direction of
the $y$-axis is large and remains almost constant during the
scattering process. For this reason we consider this motion
classically, which allows us to set $y\equiv vt$. 

In contrast, the motion along the $x$-axis is described quantum mechanically.
Moreover, we make the Raman-Nath approximation \cite{Sch,Yakovlev},
that is we neglect the kinetic energy operator
$\hat{H}_{cm}$ in the Schr\"odinger equation (\ref{Sheq}).
Hence, the displacement of the atom along the $x$-axis caused by the
interaction with the standing light wave is assumed to be small
compared to the corresponding wave length. Since $\hat{H}_{cm}$ 
is omitted in Eq. (\ref{Sheq}), the coordinate $x$ is considered 
to be a parameter. Moreover, we neglect
spontaneous emission due to the small interaction time $\tau$ and
the non-vanishing detuning $\Delta$.

As a result the Schr\"odinger equation (\ref{Sheq}) reduces to
\begin{equation}
 \label{SheqN}
    i\hbar\frac{d}{dt}|\Psi\rangle\cong\left(\hat{H}_{a}-\hat{\mathbf{d}}\cdot\mathbf{E}\right)|\Psi\rangle.
\end{equation}

In order to solve this equation we make the ansatz
\begin{equation}
 \label{wfC}
    |\Psi\rangle=A_e(t;\mathbf{r})e^{-i(\omega_e-\Delta/2)t}|e\rangle+
    A_g(t;\mathbf{r})e^{-i(\omega_g+\Delta/2)t}|g\rangle,
\end{equation}
where the amplitudes $A_g$ and $A_e$ are functions of time $t$, but
depend on the position vector $\mathbf{r}$ as a parameter. It is for
this reason that we have dropped in Eq. ({\ref{SheqN}}) the carrot
on $\mathbf{E}$.

Substituting the ansatz  Eq. (\ref{wfC}) into the approximate
Schr\"odinger equation (\ref{SheqN}), we arrive at
\begin{equation}
 \label{ampeqn}
    i\hbar\frac{d}{dt}{{A}_e\choose {A}_g}=\hat{H}{A_e\choose A_g},
\end{equation}
with the Hamiltonian
\begin{equation}
 \label{amp}
    \hat{H}\equiv\frac{1}{2}\left(\begin{array}{ccc}\hbar\Delta & V^* \\
    V & -\hbar\Delta\end{array}\right)
\end{equation}
containing the complex-valued coupling matrix element
\begin{equation}
 \label{V}
    V(t;\mathbf{r})=-2 e^{-i\omega t}\boldsymbol{\wp}\cdot\mathbf{E}(t,\mathbf{r}).
\end{equation}
Here we have introduced the dipole matrix element
$\boldsymbol{\wp}\equiv\langle g|\hat{\mathbf{d}}|e\rangle$, which can be
considered as real-valued in the case of the two-level atom.

\subsection{Matrix element of interaction}

The remaining task is to derive an explicit expression for the
matrix element $V$ defined by Eq. (\ref{V}) in the presence of the
two running waves. For this purpose we represent the sine functions
in Eq. (\ref{stwE}) as a sum of exponentials
\begin{equation}
    \mathbf{E}=\frac{\mathbf{E}_0}{2i}\left[(e^{i\mathbf{k}_2\mathbf{r}}-e^{i\mathbf{k}_1\mathbf{r}})
    e^{-i\omega t}-(e^{-i\mathbf{k}_2\mathbf{r}}-e^{-i\mathbf{k}_1\mathbf{r}})e^{i\omega t}\right]
\end{equation}
and evaluate $V$ neglecting terms oscillating with  $2\omega$ which
yields
\begin{equation}
 \label{V1}
    V(t;\mathbf{r})\simeq -i\boldsymbol{\wp}\mathbf{E}_0(\mathbf{r})
    (e^{-i\mathbf{k}_2\mathbf{r}}-e^{-i\mathbf{k}_1\mathbf{r}}).
\end{equation}

Next we substitute the explicit form of the wave vectors
$\mathbf{k}_1$ and $\mathbf{k}_2$ into Eq. (\ref{V1}) and find
\begin{equation}
 \label{V2}
    V(t;\mathbf{r})=2\boldsymbol{\wp}\mathbf{E}_0(x,y)e^{-i ky\sin\alpha}\sin(kx\cos\alpha).
\end{equation}

At this point we make use of the fact that the motion of the atom
along the $y$-axis is treated classically and we can replace the
$y$-coordinate by $y=vt$. Moreover, for the sake of simplicity we
assume that $\mathbf{E}_0$ is independent of $x$, resulting in 
$\mathbf{E}_0(x,y)=\mathbf{E}_0(vt)=\boldsymbol{\mathcal{E}}_0f(t)$, where
$f=f(t)$ denotes the envelope function along the $y$-axis, as
indicated in Fig. \ref{fig_setup}. The coupling matrix element $V$
given by (\ref{V2}) takes then the form
\begin{equation}
 \label{V4}
    V(t;\mathbf{r})=\hbar\Omega(x)f(t)e^{-i\omega_\alpha t}\;,
\end{equation}
where
\begin{equation}
 \label{Doppler}
    \omega_\alpha\equiv kv\sin\alpha
\end{equation}
and
\begin{equation}
 \label{Omega}
    \Omega(x)\equiv\frac{2}{\hbar}\boldsymbol{\wp}\boldsymbol{\mathcal{E}}_0\sin(kx\cos\alpha)
    \equiv \Omega_0\sin(kx\cos\alpha)
\end{equation}
are the velocity-dependent Doppler and position-dependent Rabi
frequencies, respectively.

Hence, the coupling matrix element $V$ consists of the product of
three terms: (i) the position-dependent coupling energy
$\hbar\Omega(x)$, (ii) the envelope function $f=f(t)$, and (iii) the
time-dependent phase factor $\exp(-i\omega_\alpha t)$ due to the
motion of the atom through the field.

\section{Dynamical and geometric phases}

In the present section we connect the Hamiltonian Eq. (\ref{amp})
together with the explicit expression  for the matrix element $V$,
Eq. (\ref{V4}), to the Hamiltonian used in Ref. \cite{Berry84} to
derive the geometric phase. This approach allows us to take
advantage of the results obtained in Ref. \cite{Berry84}. Moreover,
we establish the connection between the dressed and the atomic
states.

\subsection{Connection to Berry's approach}

For this purpose we make the identifications
\begin{equation}
 \label{R}
    X\equiv{\rm Re}\,V,\;
    Y\equiv{\rm Im}\,V,\;
    Z\equiv\hbar\Delta
\end{equation}
and the Hamiltonian (\ref{amp}) takes the form
\begin{equation}
 \label{BerryH}
    H(\boldsymbol{\mathcal{R}})=\frac{1}{2}\left(\begin{array}{ccc}
    Z & X-iY\\X+iY &-Z  \end{array}\right),
\end{equation}
where the real-valued  parameters $X$, $Y$ and $Z$ form the
cartesian coordinates of the vector $\boldsymbol{\mathcal{R}}$. Here
we have chosen the calligraphic letter rather than the normal one in
order to bring out the fact that the vector
$\boldsymbol{\mathcal{R}}$ is not in three-dimensional position
space but in parameter space.

This Hamiltonian has the eigenvalues
$\varepsilon^{(\pm)}=\pm\varepsilon$ with
\begin{equation}
 \label{epsilon}
    \varepsilon\equiv\frac{1}{2}\sqrt{X^2+Y^2+Z^2}=\frac{1}{2}\sqrt{|V|^2+(\hbar\Delta)^2}\;.
\end{equation}
and the corresponding eigenstates
\begin{equation}
 \label{eigenst}
    |\Psi^{(\pm)}\rangle\equiv\frac{1}{\sqrt{2\mathcal{R}(\mathcal{R}\pm Z)}}{Z\pm \mathcal{R}\choose X+iY},
\end{equation}
where $\mathcal{R}=|\boldsymbol{\mathcal{R}}|$.

Now we return to the determination of the adiabaticity criterion.
The condition for an adiabatic turn-on and -off of the
interaction is that at any time $t$ the rate of change of the
light-field amplitude is much smaller than the spacing between the
time-dependent quasi-energy levels. Hence, the parameters of the
system should satisfy the following inequality
\begin{equation}
 \label{ad_cr}
    \left|\frac{\hbar}{V}\frac{\partial V}{\partial t}\right|\ll|\varepsilon_+-\varepsilon_-|.
\end{equation}
The expression Eq. (\ref{V4}) for the coupling matrix element
$V$ allows us to cast Eq. (\ref{ad_cr}) into the form
\begin{equation}
 \label{adiabatic}
    \left|\frac{1}{f}\frac{df}{dt}-i\omega_\alpha\right|\ll\sqrt{\Delta^2+|\Omega|^2f^2(t)}\;.
\end{equation}
For a function $f(t)$ which decreases exponentially at large $t$, that is 
$f(t\rightarrow\pm\infty)\propto\exp(-|t|/\tau)$, we arrive at the
adiabaticity criterion
\begin{equation}
 \label{adlim}
    |\Delta|\tau\gg\sqrt{1+(\omega_\alpha\tau)^2}\;.
\end{equation}

In the adiabatic limit, we remain in the adiabatic states
$|\Psi^{(\pm)}\rangle$ which accumulate the phases in the course of
time. When the time dependence of $\boldsymbol{\mathcal{R}}$ is such
that $\boldsymbol{\mathcal{R}}(t)$ returns  at time $t=T$ to its
initial value $\boldsymbol{\mathcal{R}}(-T)$, the two instantaneous
eigenstates $|\Psi^{(\pm)}\rangle$ acquire \cite{Berry84} the
dynamical and geometric phases
\begin{equation}
 \label{gammaD}
    \varphi_D=\frac{1}{\hbar}\int\limits_{-T}^{T}\varepsilon(t)dt
\end{equation}
and
\begin{equation}
 \label{gammaG}
    \varphi_B=\iint\limits_{S_0}\frac{\boldsymbol{\mathcal{R}}\cdot d\mathbf{S}}{2\mathcal{R}^3}\;,
\end{equation}
where $S_0$ denotes the surface of integration, i.e. the surface
determined by a closed circuit forming during the one cycle of
parameter change from  $t=-T$ to $t=T$. As a result the eigenstates
after a cyclic change read
\begin{equation}
 \label{Psi-phases}
    |\Psi^{(\pm)}(T)\rangle=\exp(\mp i\varphi_D)\exp(\mp i\varphi_B)|\Psi^{(\pm)}(-T)\rangle.
\end{equation}
The geometric phase is solely determined by the flux of the
effective field $\boldsymbol{\mathcal{R}}/2\mathcal{R}^3$ through
the area enclosed by the parameter
$\boldsymbol{\mathcal{R}}=\boldsymbol{\mathcal{R}}(t)$ during one
period of the parameter change, i.e. during the time interval $2T$.

\subsection{Connection between dressed and atomic states}

The dynamical as well as the geometric phase are formulated in terms
of the dressed states Eq. (\ref{eigenst}) which arise due to the atom-field
interaction. However, in order to observe the geometric phase in
an experiment, the atom should be prepared in a well-defined free-atom
internal state. For this reason we need to connect the dressed
states $|\Psi^{(\pm)}\rangle$ with the atomic ones, that is with 
$|g\rangle$ and $|e\rangle$.

Before the atom enters the light field, that is at the time $t=-T$,
the interaction $V$ vanishes, leading to $X=Y=0$ and
$\mathcal{R}=|Z|=\hbar|\Delta|$. Here we have assumed for the sake
of simplicity that the envelope function is a mesa function with a
sharp turn-on at $-T$ and a sharp turn-off at $T$. Needless to say
this assumption is not necessary and we will consider later a smooth
envelope function. Due to appearance of the absolute value
$|\Delta|$ of the detuning it is useful to consider the two cases of
$\Delta<0$ and $\Delta>0$. Indeed, for $\Delta<0$ we find
\begin{equation}
 \label{Psi 1}
    |\Psi^{(+)}(-T)\rangle=|g\rangle\; \text{and} \; |\Psi^{(-)}(-T)\rangle=-|e\rangle,
\end{equation}
whereas for $\Delta>0$ we arrive at
\begin{equation}
 \label{Psi 2}
    |\Psi^{(+)}(-T)\rangle=|e\rangle\; \text{and} \; |\Psi^{(-)}(-T)\rangle=|g\rangle.
\end{equation}
Hence, in the case of $\Delta<0$  the ground and excited states
follow $|\Psi^{(+)}\rangle$ and $|\Psi^{(-)}\rangle$, respectively,
whereas, for $\Delta>0$ the ground and excited states follow
$|\Psi^{(-)}\rangle$ and $|\Psi^{(+)}\rangle$.

By using Eqs. (\ref{wfC}), (\ref{Psi-phases}), (\ref{Psi 1}), and (\ref{Psi 2}), 
we obtain the total phase 
\begin{equation}
 \label{g-phase}
    \varphi_g\equiv\beta\,\sign\Delta+\gamma,
\end{equation}
of the ground state acquired during the interaction time $2T$,
where 
\begin{equation}
 \label{phase-beta}
    \beta\equiv\varphi_D-|\Delta|T
\end{equation}
and
\begin{equation}
 \label{phase-gamma}
    \gamma\equiv\varphi_B\,\sign\Delta
\end{equation}
are the total dynamical and Berry phases, respectively.

The same procedure results in the total phase of the excited state
\begin{equation}
 \label{etot}
    \varphi_e=-\varphi_g-\pi\,\Theta(-\Delta),
\end{equation}
with $\Theta$ being the Heaviside function.

Note, that we have neglected the phase contributions proportional to
$\omega_g$ and $\omega_e$ arising from Eq. (\ref{wfC}), 
since they are independent of the coordinate of the atom.
Indeed, we shall show that the Berry phase as well as the dynamical phase 
depend appropriately on the position and 
therefore can be detected by a narrowing of the atomic wave packet.

\section{Circuit in parameter space}

In Sec. II we have derived the explicit expression Eq. (\ref{V4})
for the interaction matrix element $V$ of the atom-field coupling.
We are now in the position to discuss the path
$\boldsymbol{\mathcal{R}}=\boldsymbol{\mathcal{R}}(t)$ in parameter
space traversed in the course of time.

Since according to Eq. (\ref{R}) the $Z$-component of
$\boldsymbol{\mathcal{R}}$ is given by the constant detuning
$\Delta$, the circuit lies parallel to the $XY$-plane and is given
by
\begin{equation}
 \label{X}
    X(t;x)=\hbar|\Omega(x)|f(t)\cos(\omega_\alpha t)
\end{equation}
and
\begin{equation}
 \label{Y}
    Y(t;x)=-\hbar|\Omega(x)|f(t)\sin(\omega_\alpha t).
\end{equation}
Due to the position-dependence of $\Omega$ the circuit depends on
$x$ as a parameter.

This curve is most conveniently described by the polar coordinates
\begin{equation}
 \label{rho-phi}
    \rho\equiv\hbar|\Omega|f(t),\;\;\;\phi\equiv\omega_\alpha t
\end{equation}
leading to
\begin{equation}
 \label{rhophi}
    \rho=\rho(\phi)=\hbar|\Omega|f(\phi/\omega_\alpha).
\end{equation}

%%%%%%%%%%%%%%%%%%%%%%%%%%%%%%%%%%%%%%%%%%%%%%%%%%%%%%%%%%%%%%%%%%%%%%%%%%%%%%

\begin{figure}
\includegraphics[width=0.45\textwidth]{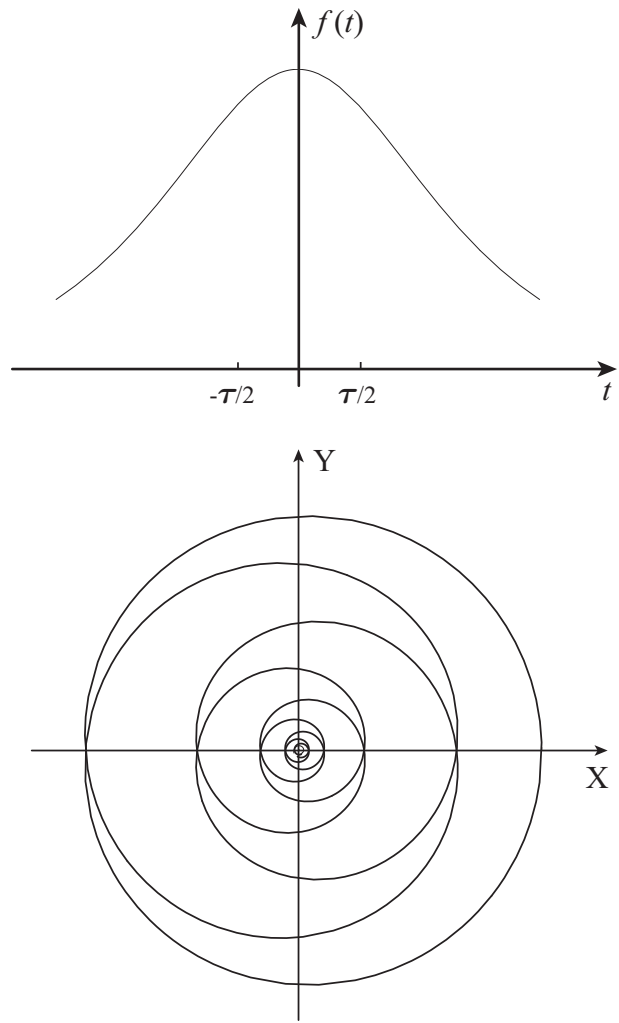}
\caption{\label{fig_circ} Translation of the electromagnetic field
envelope $f=f(t)$ shown on the top into a circuit in parameter space, 
depicted at the bottom, illustrated here for the Eckart envelope Eq. (\ref{cosh}). 
This function leads to an infinite amount of windings around 
the origin of parameter space.}
\end{figure}

%%%%%%%%%%%%%%%%%%%%%%%%%%%%%%%%%%%%%%%%%%%%%%%%%%%%%%%%%%%%%%%%%%%%%%%%%%%%%%

A model of relevant for an experiment relies on a smooth
envelope, for instance, by the Eckart envelope
\begin{equation}
 \label{cosh}
    f_E(t)\equiv \frac{1}{\cosh(t/\tau)}.
\end{equation}

Although laser beams are usually modeled by a Gaussian profile,
results obtained with the Eckart envelope are believed to be similar
to those with a Gaussian one. Moreover, in the case of the Eckart
envelope the Schr\"odinger equation (\ref{ampeqn}) for the
probability amplitudes to be in the ground and the excited state has
an exact solution, which can be used to compare the results obtained
within the Berry approach.

We emphasize that in the case of the Eckert envelope 
the amplitude $\rho$ given by Eq. (\ref{rho-phi}) 
vanishes for  $t=\pm\infty$ and consequently the
circuit starts from and terminates at the origin. In the course of
time the amplitude $\rho$ first increases and then decreases again.
At the same time the angle $\phi$ increases monotonously. As a
result the curve $\boldsymbol{\mathcal{R}}=\boldsymbol{\mathcal
R}(t)$ defining the circuit circumvents the origin infinitely many
times before it returns to it, as shown in Fig. \ref{fig_circ}. 
However, due to the increase and
decrease of $\rho$ not all windings will be visible. The number of
the prominent loops is determined by the value of
$\omega_\alpha\tau$.

\section{Explicit expressions for phases}

In the preceding section we have discussed the form of the circuit
in parameter space dictated by the longitudinal mode function of the electromagnetic field. 
In the present section we calculate the resulting geometric and dynamical phases 
and analyze the weak-field limit.

\subsection{General case}

We start by obtaining explicit formulae for the phases due to an arbitrary but smooth envelope function. 
In Appendix \ref{wkb-method} we rederive these expressions within the WKB-approach.

\subsubsection{Geometric phase}

According to Eq. ({\ref{gammaG}) we have to calculate the flux of
the effective field $\boldsymbol{\mathcal{R}}/(2\mathcal{R}^3)$
through the area enclosed by the circuit in  parameter space. Since
the vector normal to this surface is in the opposite direction to
the $Z$-axis of the Cartesian coordinate system we find that
$\boldsymbol{\mathcal{R}}\cdot d\mathbf{S}=-\hbar\Delta dS$ and Eq.
(\ref{gammaG}) with the definition Eq. (\ref{phase-gamma}) reduces to
\begin{equation}
    \gamma=-\frac{1}{2}\hbar|\Delta|
    \int dX \int dY \frac{1}{(X^2+Y^2+(\hbar\Delta)^2)^{3/2}}\;.
\end{equation}
In terms of the polar  coordinates $\rho$ and $\phi$ defined by Eq.
(\ref{rho-phi}) this integral takes the form
\begin{equation}
 \label{gamma-circuit}
    \gamma=-\frac{1}{2}\hbar|\Delta|
    \int\limits_{\phi_i}^{\phi_f}d\phi\int\limits_0^{\rho(\phi)}
    \frac{\rho\,d\rho}{(\rho^2+\hbar^2\Delta^2)^{3/2}}\;.
\end{equation}
Here the upper limit $\rho=\rho(\phi)$ of the radial integral
depends on the circuit in parameter space parameterized by the angle
$\phi$. The integration over $\phi$ runs between the initial
$\phi_i\equiv\phi(-T)$ and final $\phi_f\equiv\phi(T)$ angles of the
curve. Their precise form is dictated by the shape of the circuit.

The integration over $\rho$ can be performed and with the help of Eq. (\ref{rhophi})
and the new integration variable $t\equiv \phi/\omega_\alpha$ we arrive at
\begin{equation}
 \label{Gammaf}
    \gamma=\frac{\omega_\alpha}{2}\int\limits_{-T}^{T}dt
    \left(\frac{|\Delta|}{\sqrt{|\Delta|^2+|\Omega(x)|^2f^2(t)}}-1\right),
\end{equation}
or 
\begin{equation}
 \label{Gammaf-result}
    \gamma=\frac{\omega_\alpha}{2}\int\limits_{-T}^{T}dt
    \left(\frac{1}{\sqrt{1+a^2(x)f^2(t)}}-1\right),
\end{equation}
where
\begin{equation}
 \label{parameter-a}
    a(x)\equiv\frac{|\Omega(x)|}{|\Delta|}
\end{equation}
is the dimensionless Rabi frequency. The position dependence of the
geometric phase arises from the position dependence of $a=a(x)$.

\subsubsection{Dynamical phase}

Next we turn to the dynamical phase. For this purpose we substitute
the expression for $V$, Eq. (\ref{V4}), into the one for the
quasi-energy $\varepsilon$, Eq. (\ref{epsilon}), and find
\begin{equation}
 \label{energy}
    \varepsilon(t)=\frac{\hbar}{2}\sqrt{\Delta^2+|\Omega(x)|^2f^2(t)}.
\end{equation}

Together with Eqs. (\ref{gammaD}) and (\ref{phase-beta}) for the total dynamical phase, 
we arrive at
\begin{equation}
 \label{betaT0}
    \beta=\frac{1}{2}\int\limits_{-T}^{T}dt\left(\sqrt{|\Delta|^2+|\Omega(x)|^2f^2(t)}-|\Delta|\right),
\end{equation}
or
\begin{equation}
 \label{betaT}
    \beta=\frac{|\Delta|}{2}\int\limits_{-T}^{T}dt\left(\sqrt{1+a^2f^2(t)}-1\right),
\end{equation}
where we have recalled the definition Eq. (\ref{parameter-a}) of the dimensionless Rabi frequency $a$.

When we compare the expressions Eqs. (\ref{Gammaf}) and (\ref{betaT0}) 
for the  geometric and the dynamical phases we find the identity 
\begin{equation}
 \label{relation}
    \gamma=\omega_\alpha\frac{\partial\beta}{\partial |\Delta|}\,,
\end{equation}
which is reminiscent of the Kramers-Kronig relations. 
However, the WKB-analysis presented in Appendix A shows that 
Eq. (\ref{relation}) is merely a consequence of a Taylor 
expansion in powers of $\omega_\alpha/|\Delta|$. 
It is interesting to note that in the formalism 
employed in the preceding sections this fact is hidden.

\subsection{Weak-field limit}

Finally we consider these phases in the weak-field limit, 
that is for $a\ll 1$, when Eq. (\ref{betaT}) for the dynamical phase reduces to
\begin{equation}
 \label{beta-small-a}
    \beta\cong\frac{1}{4}\,|\Delta|a^2\int\limits_{-T}^{T}dt f^2(t),
\end{equation}
while the geometric phase given by Eq. (\ref{Gammaf-result}) reads
\begin{equation}
 \label{gamma-small-a}
    \gamma\cong -\frac{1}{4}\,\omega_\alpha a^2\int\limits_{-T}^{T}dt f^2(t).
\end{equation}
In Appendix \ref{wkb-method} we rederive these expressions directly 
by second-order perturbation theory.

A comparison between Eqs. (\ref{beta-small-a}) and (\ref{gamma-small-a}) 
reveals that in the weak-field limit, $|\Omega|\ll|\Delta|$, 
the ratio of the geometric and dynamical phases is equal to
\begin{equation}
 \label{ratio}
    \left|\frac{\gamma}{\beta}\right|=\frac{\omega_\alpha}{|\Delta|},
\end{equation}
and thus independent of the field envelope.

We conclude by estimating this ratio for typical experimental
values. For instance, for the $\quad 1s_5(J=2)\rightarrow2p_3(J=3)$
transition in argon \cite{Oberthaler}, the wave length
$\lambda=812\,\text{nm}$, the resonance detuning $\Delta\cong
3\cdot10^7\,\text{s}^{-1}$, the velocity $v_y=700\,\text{m/s}$ and
angle $\alpha=10^{-3}$, we obtain $|\gamma/\beta|\cong 0.1$, which
is feasible in an experiment.

\section{Application to Eckart envelope}

In this section we evaluate the geometric and dynamical phases 
for the Eckart envelope defined by Eq. (\ref{cosh}). 
For the details of the integrations we refer to Appendix \ref{integrals-details} and
\ref{flux-evaluation}.

\subsection{Geometric phase}

For the Eckart envelope we have $T=\infty$ and 
Eq. (\ref{Gammaf-result}) takes the form
\begin{equation}
 \label{gamma_cosh}
    \gamma_E=\omega_\alpha \tau\int\limits_0^{\infty}d\theta\left(\frac{\cosh\theta}{\sqrt{\cosh^2\theta+a^2}}-1\right),
\end{equation}
where we have introduced $\theta\equiv t/\tau$ and used the symmetry of the integrand.

In Appendix \ref{integrals-details} we perform this integral and find the geometric phase 
\begin{equation}
 \label{g_E}
    \gamma_E(x)=-\frac{1}{2}\,\omega_\alpha\tau\ln\left(1+a^2(x)\right),
\end{equation}
which in the weak-field limit $a(x)\ll 1$ reduces to
\begin{equation}
 \label{gammaE}
    \gamma_E(x)\cong -\frac{1}{2}\,\omega_\alpha\tau a^2(x).
\end{equation}

We conclude by noting that we can rederive the expression Eq.
(\ref{g_E}) for $\gamma_E$ by decomposing the path in the parameter
space into a sequence of closed circuits and calculating the sum of
the fluxes through each of these areas, as shown in Appendix
\ref{flux-evaluation}.

\subsection{Dynamical phase}

For the Eckart envelope the expression Eq. (\ref{betaT}) for the dynamical phase takes the form  
\begin{equation}
 \label{betaE1}
    \beta_E=|\Delta|\tau\int\limits_{0}^{\infty}d\theta
    \left(\sqrt{1+\frac{a^2}{\cosh^2\theta}}-1\right)
\end{equation}
and according to Appendix \ref{integrals-details} we find
\begin{equation}
 \label{betaE2}
    \beta_E=|\Delta|\tau\left[a\arctan(a)-\frac{1}{2}\ln(1+a^2)\right],
\end{equation}
which for $a^2\ll 1$ reduces to
\begin{equation}
 \label{beta-E-small a}
    \beta_E\cong\frac{1}{2}|\Delta|\tau a^2=\frac{|\Delta|}{\omega_\alpha}\,|\gamma_E|.
\end{equation}
In the last step we have recalled Eq. (\ref{gammaE}) for $\gamma_E$. 

Hence, we confirm the fact that the ratio of the geometric to the dynamical phase 
is given by Eq. (\ref{ratio}).

\section{Cancellation of dynamical phase}

Next we use the results obtained in the previous sections to
propose a scheme to cancel the dynamical phase, which always
dominates the geometric one. Moreover, due to its dependence on the
energy of the system, the dynamical phase is particularly sensitive
to the slightest change of the parameters.

From the expressions Eqs. (\ref{g-phase}) and (\ref{etot}) for the
total phases acquired by the ground and excited states, we recall
that the dynamical part given by $\beta\,\sign(\Delta)$ depends on the sign of the detuning
$\Delta$, whereas the geometric part $\gamma$ only on its absolute value.

This fact allows us to suggest a rather intuitive scheme to compensate
the dynamical phase. We propose to use two consecutive interactions of the
atom with the standing waves, that is, firstly with blue-detuned
waves ($\Delta>0$), secondly with red-detuned waves ($\Delta'\equiv
-\Delta<0$). Here we have introduced a prime to indicate the second
standing wave. As a result of the opposite signs of the
detunings, the dynamical contributions cancel each other provided 
the condition $k\cos\alpha=k'\cos\alpha'$ is fulfilled, or when the
position-dependent Rabi frequencies $\Omega(x)$ and $\Omega'(x)$
defined by Eq. (\ref{Omega}) are equal. 

At the same time, the geometric phases add up and result in the total phase 
\begin{equation}
    \varphi_g^{(tot)}=\varphi_g+\varphi_g'=\gamma+\gamma'
\end{equation}
of the ground state.

Thus, in a scheme of two consecutive scatterings of the atom by
oppositely detuned standing waves, we obtain a cancellation of the
dynamical phase and summation of the Berry phase. Of course, 
the time interval between the two interaction zones should
be larger than the interaction time $\tau$ itself, in order to be
consistent with the adiabatic approximation on the turn-on and
turn-off of the interactions.

\section{Focusing due to geometric phase}

In the previous section we have shown that in our scattering setup 
with first an interaction with the red-detuned wave, 
and then with the blue-detuned wave, the atom
acquires only the geometric phase. We now demonstrate that this phase
manifests itself in a focusing of the atomic wave packet.

For this purpose we assume for the wave function of the
center-of-mass motion of an atom in the ground state moving
in the $y$-direction a Gaussian 
\begin{equation}
 \label{Psi0i}
    \Psi_0(x)\equiv\frac{1}{(\sqrt{\pi}\Delta x_0)^{1/2}}
    \exp\left[-\frac{x^2}{2\Delta x_0^2}+i\gamma(x)\right]
\end{equation}
of width $\Delta x_0$ and the additional scattering-induced
geometric phase $\gamma(x)$.

The time evolution of this wave packet in the absence of an external field
is given by
\begin{equation}
 \label{Psip}
    \Psi(x,t)=\sqrt{\frac{M}{2\pi i\hbar t}}\int\limits_{-\infty}^{\infty}dx'
    \exp\left[i\frac{M}{2\hbar\,t}(x-x')^2\right]\Psi_0(x').
\end{equation}

In the case of the Eckart envelope  and in the limit of 
$k\Delta x_0<1$ and $(|\Omega_0|/|\Delta|)(k\Delta x_0)<1$, 
the geometric phase $\gamma(x)$ given by Eq. (\ref{g_E}) near 
the nodes of the standing wave, for instance near $x=0$, 
is a quadratic function 
\begin{equation}
 \label{B_lens}
    \gamma(x)\cong -\frac{b}{2}\frac{x^2}{\Delta x_0^2}
\end{equation}
of $x$ with
\begin{equation}
 \label{b}
    b\equiv \omega_\alpha\tau\,\frac{|\Omega_0|^2}{|\Delta|^2}(k\cos\alpha\Delta x_0)^2.
\end{equation}
Here we have used the definitions Eqs. (\ref{Omega}) and (\ref{parameter-a}).

By substituting the initial wave function Eq. (\ref{Psi0i}) into 
Eq. (\ref{Psip}), we obtain the time-dependent distribution
\begin{equation}
    \left|\Psi(x,t)\right|^2=\frac{1}{\sqrt{\pi}
    \Delta x(t)}\exp\left(-\frac{x^2}{\Delta x^2(t)}\right)
\end{equation}
of finding an atom with the coordinate $x$, where the width  
\begin{equation}
 \label{wp-berry}
    \Delta x(t)\equiv\Delta x_0\sqrt{\left(1-b\frac{t}{t_{s}}\right)^2+\left(\frac{t}{t_{s}}\right)^2}.
\end{equation}
is determined by the initial width $\Delta x_0$ and the Berry phase contribution.
Here $t_{s}\equiv M\Delta x_0^2/\hbar$ is the characteristic time of field-free spreading, 
that is 
\begin{equation}
 \label{x0}
    \Delta x^{(0)}(t)\equiv\Delta x_0\sqrt{1+\frac{t^2}{t_{s}^2}}\,.
\end{equation}

The minimal possible width 
\begin{equation}
 \label{xmin}
    \Delta x_{min}\equiv\frac{\Delta x_0}{\sqrt{1+b^2}}
\end{equation}
of the packet is reached at the time
\begin{equation}
    t_{min}=\frac{b}{1+b^2}\,t_{s}.
\end{equation}

%%%%%%%%%%%%%%%%%%%%%%%%%%%%%%%%%%%%%%%%%%%%%%%%%%%%%%%%%%%%%%%%%%%%%%%%%%%%%%

\begin{figure}
\includegraphics[width=0.45\textwidth]{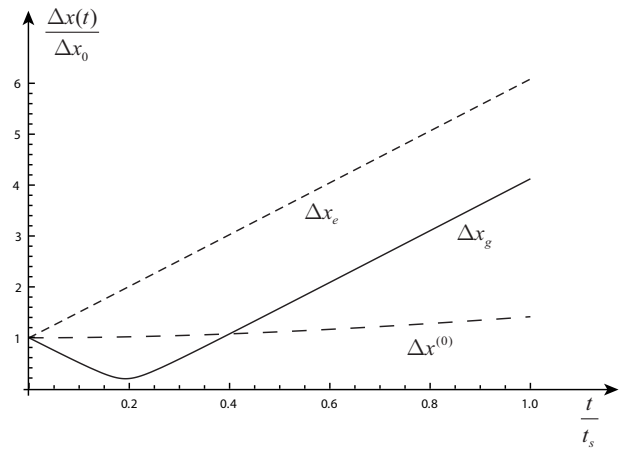}
\caption{Influence of the geometric phase on the free propagation 
of an atomic wave packet represented by its time-dependent width $\Delta x(t)$ 
given by Eq. (\ref{wp-berry}). For an atom in the ground state ($\Delta x_g$ and $b=5$) 
the wave packet first focuses and then spreads, whereas 
for the excited state ($\Delta x_e$ and $b=-5$) it spreads from the beginning. 
This spreading is larger than that associated with the free propagation of the Gaussian 
($\Delta x^{(0)}$ and $b=0$) given by Eq. (\ref{x0}).}
\end{figure}

%%%%%%%%%%%%%%%%%%%%%%%%%%%%%%%%%%%%%%%%%%%%%%%%%%%%%%%%%%%%%%%%%%%%%%%%%%%%%%

In Fig. 3, we present the focusing of the atomic wave packet
induced by the geometric phase for relatively small values of the
parameter $b$. For example, the value $b\approx 5$ is achieved for 
$\omega_\alpha\tau=4\pi$, the Rabi frequency $\Omega_0=1.8\,|\Delta|$ 
and the initial width $k\Delta x_0=0.25$. For $b=5$ the expression Eq. (\ref{xmin}) 
for the minimal width predicts a shrinking up to around
five times compared to the initial width, as indicated by 
the solid curve on Fig. 3 denoted by $\Delta x_g$. 
After the point $t\approx 0.19\,t_{s}$ the wave packet expands faster than the free wave packet
represented by the dashed line $\Delta x^{(0)}$. 

The atom in the excited state acquires the same geometric phase as the ground state but
with the opposite sign. Therefore, the parameter $b$ determining the narrowing appears
with the opposite sign and results in an accelerated spreading rather than a focusing of
the wave packet, as shown by $\Delta x_e$ in Fig. 3.

\section{Conclusions}

In this article we propose a scheme to observe the geometric phase
in atom optics based on the scattering of a two-level atom by two
consecutive standing light waves with the same envelope but
opposite detunings.
The dynamical and the geometric phases acquired by the two-level
atom during the interaction are calculated within 
(i) the rotating wave approximation, (ii) the Raman-Nath approximation, 
and (iii) adiabatically slow switch-on and -off of the interaction. 
Both the dynamical and the geometric phases are evaluated for 
a field envelope given by the Eckart function.

We now specify the conditions under which these approximations 
are valid and self-consistent. Indeed, the velocity acquired due to a
resonant atom-field interaction can be estimated as
$$
\langle v_x \rangle\sim\frac{\hbar k}{M}\,\Omega_0\tau,
$$
where $\Omega_0$ is a characteristic value of the Rabi frequency.
Hence, we can omit the kinetic energy operator in the Schr\"odinger
equation when $ \langle v_x\rangle\tau\ll1/k$, or
$$
\omega_{rec}\tau^2\Omega_0\ll 1,
$$
with $ \omega_{rec}\equiv\hbar k^2/(2M)$ being the recoil frequency.

Moreover, we can neglect the spontaneous emission due to the small
interaction time $\tau$, provided
$$
\Gamma\tau w\lesssim 1,
$$
where $\Gamma$ and $w$ are the spontaneous emission rate and
the occupation probability of the excited level, respectively. In the
case of near resonance, $|\Delta|>\Omega_0$, the maximum value of
the population probability is $w\sim(\Omega_0/\Delta)^2$.

In a sequence of two such scattering arrangements with opposite signs
of their detunings the dynamical phases compensate each other, 
whereas the geometric phases add. Therefore, the final state of
the atom is different from the initial one only by the acquired Berry
phase, which in our case depends not only on the internal, but also
on the external degrees of freedom, such as the position of atom.
This dynamical phase cancellation provides us with 
a completely different technique of measuring the geometric phase, 
employing the narrowing of the atomic wave-packet prepared
initially in the ground state. This novel suggestion of the Berry
phase measurement based on self-lensing beneficially differs from
the previous interferometric schemes of observing the geometric
phase and might also constitute a useful tool in atomic lithography.

The familiar WKB-technique is used as an independent method to
verify the results obtained within the standard approach
\cite{Berry84} and the results are shown to coincide. 
However, we emphasize that the treatment of Ref. \cite{Berry84} 
is more general in the sense that it can be employed for any dependence 
of the interaction envelope and the detuning on time. 
We consider in this paper the particular case of a constant detuning.    

We conclude by emphasizing again that the Raman-Nath regime 
takes place for a small interaction time $\tau$, that is $\tau\ll
1/(\omega_{rec}\Omega_0)^{1/2}$. In this case, we are allowed to
neglect the kinetic energy operator. However, for large interaction
times, that is for $1/(\omega_{rec}\Omega_0)^{1/2}\ll\tau\ll 1/\Gamma$ 
it should be taken into account. To do this in an exact way we could 
consider the scattering of the atom by a sequence of two near-resonant 
{\it running} rather than standing light waves \cite{PM}. 
In this case the formulae for the geometric and dynamical phases 
are analogous to those of the standing wave case, but independent on 
the external atomic degrees of freedom, i.e. the atomic position. 
Therefore, no lensing occurs and we can only use the standard 
interferometric scheme for the observation of the geometric phase.

\section*{Acknowledgments}

We are deeply indebted to J. Baudon, M.V. Fedorov, and V.P. Yakovlev 
for many suggestions and stimulating discussions. MAE is
grateful to the support of the Alexander von Humboldt Stiftung and 
the Russian Foundation for Basic Research (grant 10-02-00914-a). 
PVM acknowledges support from the EU project "CONQUEST" and the German Academic
Exchange Service.

\begin{appendix}
\section{Alternative derivation of phases}

\label{wkb-method}

In this appendix we present alternative routes toward the expressions 
Eqs. (\ref{Gammaf-result}) and (\ref{betaT}) obtained within Berry's formalism.
To solve the Schr\"odinger equation (\ref{SheqN}) we make the ansatz
\begin{equation}
 \label{wfCA}
    |\Psi\rangle=\tilde{A}_e(t;{\bf r})\,e^{-i(\omega_e-\tilde{\Delta}/2)t}|e\rangle+
    \tilde{A}_g(t;{\bf r})\,e^{-i(\omega_g+\tilde{\Delta}/2)t}|g\rangle,
\end{equation}
which is different from Eq. (\ref{wfC}) due to the effective
detuning
\begin{equation}
 \label{delta-eff}
    \tilde{\Delta}\equiv\Delta+\omega_\alpha.
\end{equation}

Substituting Eq. (\ref{wfCA}) into Eq. (\ref{SheqN}) and making the
rotating wave approximation, we arrive at
\begin{equation}
 \label{ampeqnA}
 i\hbar\frac{d}{d t}\left({\tilde{A}_e \atop \tilde{A}_g}\right)=
 \frac{1}{2}
 \begin{pmatrix} \hbar\tilde{\Delta} & \tilde{V}\\ \tilde{V} &-\hbar\tilde{\Delta}\end{pmatrix}
 \left({\tilde{A}_e \atop \tilde{A}_g}\right),
\end{equation}
where in contrast to $V$ given by Eq. (\ref{V4}) the coupling matrix element
\begin{equation}
 \label{V-real}
    \tilde{V}(t;\mathbf{r})=\hbar\Omega(x)f(t)\equiv\hbar\tilde{\Omega}(x,t)
\end{equation}
is now real and depends on time only through the envelope $f(t)$,
describing the adiabatic turn-on and turn-off of the interaction.

Two approaches to solve Eq. (\ref{ampeqnA}) and obtain the relevant phases 
offer themselves : (i) the WKB-technique and (ii) second-order perturbation theory.

\subsection{WKB-approach}

In order to transform the two first order differential equations, Eq. (\ref{ampeqnA}), 
into a single one of second order, we introduce the two functions
\begin{equation}
 \label{uvA}
    \left({u}\atop
    {v}\right)\equiv\frac{1}{2}\left({\tilde{A}_g+\tilde{A}_e}\atop{\tilde{A}_e-\tilde{A}_g}\right),
\end{equation}
resulting in the differential equation \cite{Fedorov}
\begin{equation}
 \label{equ}
    \frac{d^2}{dt^2}u+\frac{1}{4}\left(\tilde{\Delta}^2+\tilde{\Omega}^2+2i\frac{d\tilde{\Omega}}{dt}\right)u=0
\end{equation}
for $u=u(t)$.

This equation can be represented in a form similar to the stationary
Schr\"{o}dinger equation in position space
by introducing the dimensionless variable $\xi\equiv t/\tau$,
where $\tau$ is the characteristic time
scale of the envelope $f$. In terms of $\xi$, Eq. (\ref{equ})
reads
\begin{equation}
 \label{uE}
    \frac{1}{\tau^2}\frac{d^2}{d\xi^2}\;u+
    \left(\tilde{\varepsilon}^2+\frac{i}{2\tau}\frac{d\tilde{\Omega}}{d\xi}\right)u=0,
\end{equation}
where
\begin{equation}
 \label{E}
    \tilde{\varepsilon}(\xi)\equiv\frac{1}{2}\sqrt{\tilde{\Delta}^2+\tilde{\Omega}^2(\xi)}\;.
\end{equation}

Equation (\ref{uE}) is analogous to the stationary
one-dimensional Schr\"odinger equation where $\xi$ and
$\tilde{\varepsilon}$ play the role of the "effective coordinate"
and "effective momentum", respectively. The small parameter $1/\tau$ mimics 
Planck's constant in the conventional WKB-approach.

Due to the adiabaticity condition
\begin{equation}
    \frac{1}{\tau\tilde{\varepsilon}^2}\cfrac{d\tilde{\varepsilon}}{d\xi}\ll 1,
\end{equation}
we can employ the semiclassical approach to search 
for the solution $u(\xi)$  in the form
\begin{equation}
 \label{wkb-u}
    u(\xi)=e^{i\tau S(\xi)},
\end{equation}
where the complex-valued function $S(\xi)$ obeys the equation
\begin{equation}
 \label{wkb-equation}
    -\left(\frac{dS}{d\xi}\right)^2+\tilde{\varepsilon}^2+
    \frac{i}{\tau}\frac{d^2 S}{d\xi^2}+\frac{i}{2\tau}\frac{d\tilde{\Omega}}{d\xi}=0\,.
\end{equation}

Within the WKB approach $S(\xi)$ is expanded into the perturbation series 
\begin{equation}
 \label{wkb-expension}
    S(\xi)=S^{(0)}(\xi)+\frac{1}{\tau}S^{(1)}(\xi)+\frac{1}{\tau^2}S^{(2)}(\xi)+\,...
\end{equation}
in powers of $1/\tau$ and from Eq. (\ref{wkb-equation}) 
we find the zero-order term 
\begin{equation}
 \label{wkb-zero}
    S^{(0)}(\xi)=\pm\int\limits_{-T/\tau}^{\xi}d\xi'\;\tilde{\varepsilon}(\xi')
\end{equation}
and the first adiabatic correction
\begin{equation}
 \label{wkb-first}
    S^{(1)}(\xi)=\frac{i}{2}
    \left[\ln\left(\frac{2\tilde{\varepsilon}(\xi)}{|\tilde{\Delta}|}\right)\pm\ln{\mathcal N}(\xi)\right]
\end{equation}
with 
\begin{equation}
 \label{wkb-N}
    {\mathcal N}(\xi)\equiv\frac{\tilde{\Omega}(\xi)}{|\tilde{\Delta}|}+
      \left(1+\frac{\tilde{\Omega}^2(\xi)}{\tilde{\Delta}^2}\right)^{1/2}.
\end{equation}

The second term on the right-hand 
side of Eq. (\ref{wkb-first}) is a consequence of the last term in 
the left-hand side of Eq. (\ref{wkb-equation}) 
and appears in addition to the conventional WKB solution.   
Moreover, the next-order correction to $S$
gives negligible contribution to $u$. 

The general solution of Eq. (\ref{uE}) can then be written in the form
\begin{equation}
 \label{ut}
    u(\xi)=\sqrt{\frac{|\tilde{\Delta}|}{2\tilde{\varepsilon}}}\sum_{\pm}A_{\pm}
    \exp\left[\pm i\tau\int\limits_{-T/\tau}^{\xi}d\xi'\tilde{\varepsilon}(\xi')\mp \frac{\ln{\mathcal N(\xi)}}{2}\right],
\end{equation}
where the coefficients $A_{\pm}$ can be found from the initial conditions
$\tilde{A}_g (-T)=1$ and $\tilde{A}_e (-T)=0$ for 
$\tilde{A}_{g}(t)$ and $\tilde{A}_{e}(t)$. By using Eq. (\ref{uvA}) 
and the connection between $u$ and $v$ we get $u(\xi)=1/2$ and 
$(du/d\xi)=i\tilde{\Delta}\tau/4$ at $\xi=-T/\tau$, which
results in
\begin{equation}
 \label{AB}
    A_{\pm}=\frac{1}{4}(1\pm\sign\tilde{\Delta})=\frac{1}{2}\,\Theta(\pm\tilde{\Delta}).
\end{equation}
Here we have used the fact that at $\xi=-T/\tau$ 
the Rabi frequency $\tilde{\Omega}$  defined by Eq. (\ref{V-real}) vanishes and therefore
${\mathcal N}(-T/\tau)=1$ and $2\tilde{\varepsilon}(-T/\tau)=|\tilde{\Delta}|$.

The same procedure can be applied to find $v(t)$,
leading to $v(t)=-u(t)$. We then obtain the relations
$\tilde{A}_g=u-v=2u$ and $\tilde{A}_e=u+v=0$. 
The latter confirms the fact that there are no transitions 
into the excited state due to the adiabatically slow time dependence of the interaction.

Thus, the probability amplitude $\tilde{A}_g$ is given by
\begin{equation}
 \label{wkb-solution}
    \tilde{A}_g(t)=\sqrt{\frac{|\tilde{\Delta}|}{2\tilde{\varepsilon}}}
    \sum_{\pm}\Theta(\pm\tilde{\Delta})
    \exp\left[\pm i\tau\int\limits_{-T}^{t}dt'\tilde{\varepsilon}(t')\mp\frac{\ln{\mathcal N}}{2}\right].
\end{equation}

By taking into account the exponential prefactor
$\exp(-i\tilde{\Delta}t/2)$ in Eq. (\ref{wfCA}), we obtain from Eq.
(\ref{wkb-solution}) the total phase 
\begin{equation}
 \label{phaseWKB}
    \tilde{\varphi}_g=\frac{\sign(\tilde{\Delta})}{2}
    \int\limits_{-T}^{T}dt\left(\sqrt{\tilde{\Delta}^2+\tilde{\Omega}^2(t)}-|\tilde{\Delta}|\right)
\end{equation}
acquired by the ground state at $t=T$, when the interaction switches off, 
that is $\tilde{A}_g(T)\equiv\exp(i\tilde{\varphi}_g)$. 

According to Eq. (\ref{phaseWKB}) the total phase 
$\tilde{\varphi}_g$ is a function of the effective detuning $\tilde{\Delta}=\Delta+\omega_\alpha$.
When $\omega_{\alpha}\ll |\Delta|$ we can expand $\tilde{\varphi}_g$
into a Taylor series 
\begin{equation}
 \label{phaseWKB-expension}
    \tilde{\varphi}_g(\tilde{\Delta})=\tilde{\varphi}_g(\Delta+\omega_\alpha)\cong \tilde{\varphi}_g(\Delta)+
    \omega_\alpha\frac{\partial\tilde{\varphi}_g}{\partial\Delta}
\end{equation}
over $\omega_\alpha/|\Delta|$ and arrive at
$$
\tilde{\varphi}_g=\frac{\Delta}{2}\int\limits_{-T}^{T}dt\left(\sqrt{1+a^2f^2(t)}-1\right)
$$
\begin{equation}
 \label{phaseWKB-integral}
    +\frac{\omega_\alpha}{2}\int\limits_{-T}^{T}dt\left(\frac{1}{\sqrt{1+a^2f^2(t)}}-1\right).
\end{equation}
Here we have recalled the definitions Eqs. (\ref{parameter-a}) and (\ref{V-real}). 

A comparison between Eqs. (\ref{g-phase}) and (\ref{phaseWKB-integral}) reveals that 
the first term in Eq. (\ref{phaseWKB-integral}) gives the expression Eq. (\ref{betaT}) 
for the dynamical phase, whereas the second term is the geometric phase given by Eq. (\ref{Gammaf-result}). 
Moreover, Eq. (\ref{phaseWKB-expension}) shows that the Kramers-Kronig-like 
relation, Eq. (\ref{relation}), between the dynamical and
geometric phases is a consequaence of a Taylor expension.  

We conclude by noting that Eq. (\ref{equ}) for the Eckart envelope
can be solved exactly \cite{Fedorov} in terms of hypergeometric functions.
In the adiabatic limit, $|\Delta|\tau\gg 1$ and
$\omega_{\alpha}\ll |\Delta|$, the probability amplitudes
$\tilde{A}_{g}$ and $\tilde{A}_{e}$ obtained from the exact solution
coincide \cite{PM} with those derived within the WKB approach.

\subsection{Perturbation theory}

The Schr\"odinger equation (\ref{ampeqnA}) can be solved
perturbatively  using the coupling matrix element $\tilde{V}$ as the expansion parameter. 
Indeed, the second-order correction to the probability
amplitudes $\tilde{a}_e\equiv\tilde{A}_e\,e^{i\tilde{\Delta}t/2}$ and
$\tilde{a}_g\equiv\tilde{A}_g\,e^{-i\tilde{\Delta}t/2}$ are given by
\begin{equation}
 \label{Ae-corrections}
    \tilde{a}_e^{(2)}(T)=-\frac{\tilde{a}_e(-T)}{4}
    \int\limits_{-T}^{T}dt\tilde{V}(t)e^{i\tilde{\Delta}t}
    \int\limits_{-T}^{t}dt'\tilde{V}(t')e^{-i\tilde{\Delta}t'}
\end{equation}
and
\begin{equation}
 \label{Ag-corrections}
    \tilde{a}_g^{(2)}(T)=-\frac{\tilde{a}_g(-T)}{4}
    \int\limits_{-T}^{T}dt\tilde{V}(t)e^{-i\tilde{\Delta}t}
    \int\limits_{-T}^{t}dt'\tilde{V}(t')e^{i\tilde{\Delta}t'}.
\end{equation}

By using the adiabaticity condition Eq. (\ref{ad_cr}), we find
$$
\int\limits_{-T}^{t}dt'\tilde{V}(t')e^{\pm i\tilde{\Delta}t'}\cong
\mp\,\frac{i\tilde{V}(t)}{\tilde{\Delta}}\,e^{\pm i\tilde{\Delta}t}\,,
$$
which results in
$$
\tilde{a}_e(T)=\tilde{a}_e(-T)+\tilde{a}_e^{(2)}=
\tilde{a}_e(-T)\left[1-\frac{i}{4}\int\limits_{-T}^{T}dt\frac{\tilde{V}^2(t)}{\tilde{\Delta}}\right],
$$
that is
\begin{equation}
 \label{Ae-corrections-result}
    \tilde{a}_e(T)\cong\tilde{a}_e(-T)e^{-i\varphi}
\end{equation}
with the phase
\begin{equation}
 \label{phase-appendix}
   \varphi\equiv\frac{1}{4}\int\limits_{-T}^{T}dt\frac{\tilde{V}^2(t)}{\tilde{\Delta}}.
\end{equation}

Similarly, we obtain
$$
\tilde{a}_g(T)=\tilde{a}_g(-T)+\tilde{a}_g^{(2)}=\tilde{a}_g(-T)
\left[1+\frac{i}{4}\int\limits_{-T}^{T}dt\frac{\tilde{V}^2(t)}{\tilde{\Delta}}\right],
$$
or
\begin{equation}
 \label{Ag-corrections-result}
    \tilde{a}_g(T)\cong\tilde{a}_g(-T)e^{i\varphi}.
\end{equation}

By using the definitions Eqs. (\ref{parameter-a}), (\ref{delta-eff}) and (\ref{V-real}) 
we derive for the total phase $\varphi$ given by Eq. (\ref{phase-appendix}) the expression  
\begin{equation}
    \varphi=\frac{a^2}{4}\frac{\Delta^2}{\Delta+\omega_\alpha}\int\limits_{-T}^{T}dtf^2(t).
\end{equation}
In the case when $\omega_\alpha\ll|\Delta|$ the phase reads
\begin{equation}
 \label{phase-perturbation}
    \varphi\cong\frac{a^2}{4}\Delta\int\limits_{-T}^{T}dtf^2(t)-\frac{a^2}{4}\omega_\alpha\int\limits_{-T}^{T}dtf^2(t).
\end{equation}
With the help of Eq. (\ref{g-phase}) we find that the first and second terms in Eq. (\ref{phase-perturbation}) 
give the dynamical and geometric phases in the weak-field limit 
defined by Eqs. (\ref{beta-small-a}) and (\ref{gamma-small-a}), respectively.

\section{Evaluation of integrals}

\label{integrals-details}

In this appendix we evaluate the integrals Eqs. (\ref{gamma_cosh}) and (\ref{betaE1}) 
determining the geometric and dynamical phase in the case of the Eckart envelope. 
We note that some of the relations are also useful for Appendix \ref{flux-evaluation}.

\subsection{Integral determining the geometric phase}

In order to calculate the integral
\begin{equation}
    I_E\equiv\int\limits_0^{\infty}d\theta\left[\frac{\cosh\theta}{\sqrt{\cosh^2\theta+a^2}}-1\right]
\end{equation}
we recall the integral relation
\begin{equation}
 \label{gamma-integral}
    \int d\theta\frac{\cosh\theta}{\sqrt{\cosh^2\theta+a^2}}=
    \ln\left(\sinh\theta+\sqrt{\cosh^2\theta+a^2}\;\right)
\end{equation}
and find
\begin{equation}
    I_E=\lim_{\theta\rightarrow\infty}
    \left\{\ln\left[\frac{\sinh\theta+\sqrt{\cosh^2\theta+a^2}}{\sqrt{1+a^2}}\right]-\theta\right\},
\end{equation}
which with the help of the asymptotic behavior
\begin{equation}
 \label{sin-cos}
    \sinh\theta\cong\cosh\theta\cong\frac{1}{2}\,e^{\theta}
\end{equation}
in the limit of $\theta\rightarrow\infty$ reduces to
\begin{equation}
    I_E=-\frac{1}{2}\ln\left(1+a^2\right).
\end{equation}

\subsection{Integral determining the dynamical phase}

The dynamical phase is determined by the integral
\begin{equation}
    \tilde{I}_E\equiv \int\limits_{0}^{\infty}d\theta\left(\sqrt{1+\frac{a^2}{\cosh^2\theta}}-1\right),
\end{equation}
or
$$
\tilde{I}_E\equiv
a^2\int\limits_{0}^{\infty}\frac{d\theta}{\cosh^2\theta}\frac{1}{1+\sqrt{1+a^2\cosh^{-2}\theta}}.
$$
The change of the integration variable
$$
\sin\vartheta\equiv\sqrt{\frac{a^2}{1+a^2}}\,\tanh\theta
$$
gives
\begin{equation}
    \tilde{I}_E=a\left[\arctan(a)-\int\limits_{0}^{\arctan(a)}\frac{d\vartheta}{1+\sqrt{1+a^2}\cos\vartheta}\right].
\end{equation}
When we recall the integral relation
$$
\int\limits_{0}^{\vartheta_0}\frac{d\vartheta}{A+B\cos\vartheta}=
$$
$$
\frac{1}{\sqrt{B^2-A^2}}
\ln\Bigg|\frac{\sqrt{B^2-A^2}\tan(\vartheta_0/2)+A+B}{\sqrt{B^2-A^2}\tan(\vartheta_0/2)-A-B}\Bigg|
$$
for $A<B$, we arrive at
\begin{equation}
    \tilde{I}_E=a\arctan(a)-\frac{1}{2}\ln(1+a^2).
\end{equation}

\section{Evaluation of flux}
\label{flux-evaluation}

In this appendix we evaluate the flux through the area in parameter
space defined by the trajectory following from the Eckart envelope
of the electromagnetic field. In this case the separation of 
the trajectory from the origin, determined by
the strength of the envelope, first increases from zero to a maximum
value and then decreases again to zero. As a result, we obtain
infinitely many windings and thus infinitely many areas of different sizes.

%%%%%%%%%%%%%%%%%%%%%%%%%%%%%%%%%%%%%%%%%%%%%%%%%%%%%%%%%%%%%%%%%%%%%%%%%%%%%%

\begin{figure}
\includegraphics[width=0.45\textwidth]{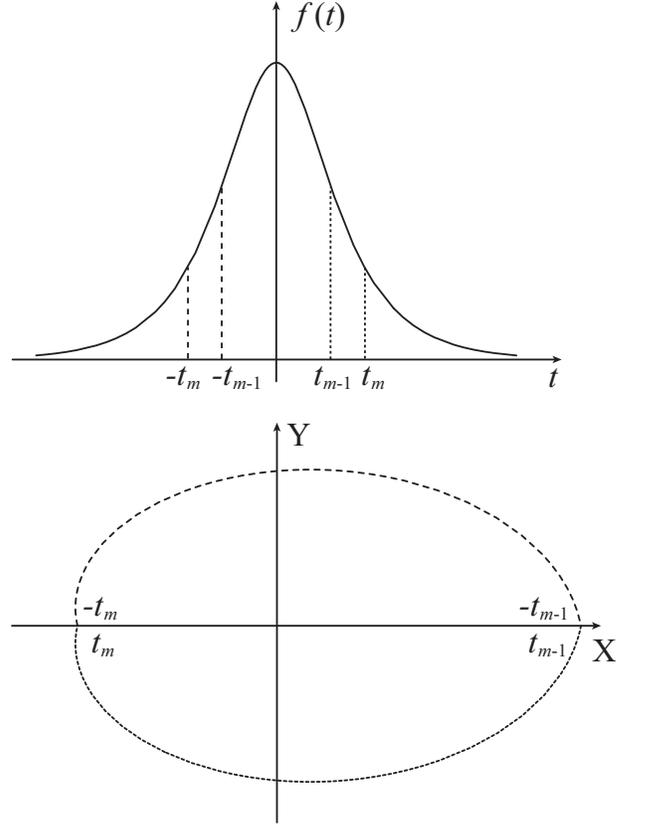}
\caption{\label{fig_COSH} Translation of the field envelope $f=f(t)$
during two symmetrically located time intervals (top) into a closed
circuit in parameter space (bottom). Indeed, the $m$-th  circuit
arises from the envelope $f=f(t)$ during the time intervals
$-t_m\leq t\leq-t_{m-1}$ and $t_{m-1}\leq t\leq t_{m}$.}
\end{figure}

%%%%%%%%%%%%%%%%%%%%%%%%%%%%%%%%%%%%%%%%%%%%%%%%%%%%%%%%%%%%%%%%%%%%%%%%%%%%%%

In order to calculate the total flux through them, we present the expression Eq. (\ref{gamma-circuit})  
for the geometric phase $\gamma_E$ as a sum 
\begin{equation}
 \label{gamma-phi}
    \gamma=-\sum_{m}\frac{1}{2}\int\limits_{\phi_i^{m}}^{\phi_f^{m}}d\phi
    \int\limits_{0}^{\rho(\phi)}\frac{\tilde{\rho}\,d\tilde{\rho}}{(1+\tilde{\rho}^2)^{3/2}}
    \equiv\sum_{m}\gamma^{(m)}
\end{equation}
of the fluxes through the $m$-th area with
\begin{equation}
 \label{gamma-rho}
    \rho(\phi)\equiv\frac{a}{\cosh[\phi/(\omega_\alpha\tau)]}.
\end{equation}

Here we have recalled the definitions Eqs. (\ref{rhophi}), (\ref{cosh}) and (\ref{parameter-a}).

In order to perform the integration in Eq. (\ref{gamma-phi}) we need to
determine the angles $\phi$ corresponding to the path defining the
$m$-th area. This path is dictated by the envelope during the time
intervals $-t_m\leq t\leq -t_{m-1}$ and $t_{m-1}\leq t\leq t_{m}$,
where $t_m\equiv m\pi/\omega_{\alpha}$, as shown in Fig.
\ref{fig_COSH}. Since $\phi=\omega_\alpha t$ these time domains
translate into the integration intervals $-\phi_m\leq \phi\leq
-\phi_{m-1}$ and $\phi_{m-1}\leq \phi\leq \phi_{m}$ with
$\phi_m\equiv m\pi$. 

Therefore, the geometric phase $\gamma^{(m)}$
given by the flux through the $m$-th area reads
$$
\gamma^{(m)}=-\frac{1}{2}\int\limits_{-m\pi}^{-(m-1)\pi}d\phi
 \int\limits_{0}^{\rho(\phi)}\frac{\tilde{\rho}\,d\tilde{\rho}}{(1+\tilde{\rho}^2)^{3/2}}
$$
\begin{equation}
    -\frac{1}{2}\int\limits_{(m-1)\pi}^{m\pi}d\phi
     \int\limits_{0}^{\rho(\phi)}\frac{\tilde{\rho}\,d\tilde{\rho}}{(1+\tilde{\rho}^2)^{3/2}}\,,
\end{equation}
which due to the symmetry of the Eckart envelope giving rise via Eq. (\ref{gamma-rho}) to $\rho(-\phi)=\rho(\phi)$  
simplifies to 
\begin{equation}
    \gamma^{(m)}=-\int\limits_{(m-1)\pi}^{m\pi}d\phi
     \int\limits_{0}^{\rho(\phi)}\frac{\tilde{\rho}\,d\tilde{\rho}}{(1+\tilde{\rho}^2)^{3/2}}\,,
\end{equation}
or
\begin{equation}
 \label{gm-integral}
    \gamma^{(m)}=\omega_\alpha\tau\int\limits_{\theta_{m-1}}^{\theta_m}d\theta
\left(\frac{\cosh\theta}{\sqrt{\cosh^2\theta+a^2}}-1\right)
\end{equation}
with $\theta_m\equiv m\pi/(\omega_\alpha\tau)$.

With the help of the integral relation Eq. (\ref{gamma-integral}) we
arrive at
\begin{equation}
 \label{gm}
    \gamma^{(m)}=\omega_\alpha\tau\left(\ln\frac{F_m}{F_{m-1}}-\frac{\pi}{\omega_\alpha\tau}\right),
\end{equation}
where 
$$
F_m\equiv\sinh\theta_m+\sqrt{\cosh^2\theta_m+a^2}.
$$

The total geometric phase $\gamma_E$ is the sum of the fluxes through all areas, that is
\begin{equation}
 \label{gsum}
    \gamma_E=\sum_{m=1}^{\infty}\gamma^{(m)}=\omega_\alpha\tau
\lim_{N\rightarrow\infty}\sum_{m=1}^{N}\left(\ln\frac{F_m}{F_{m-1}}-\frac{\pi}{\omega_\alpha\tau}\right).
\end{equation}
Using the functional relation
$$
\ln\frac{F_m}{F_{m-1}}=\ln F_m-\ln F_{m-1}
$$
of the logarithm we find
\begin{equation}
    \gamma_E=\omega_\alpha\tau\lim_{N\rightarrow\infty}
    \left(\ln F_N- \ln F_{0}-N\frac{\pi}{\omega_\alpha\tau}\right).
\end{equation}
The asymptotic expansion Eq. (\ref{sin-cos}) yields
\begin{equation}
    F_N\cong\exp\left(N\frac{\pi}{\omega_\alpha\tau}\right),
\end{equation}
which together with $F_0=(1+a^2)^{1/2}$ leads us to the total
geometrical phase
\begin{equation}
 \label{gamE}
    \gamma_E=-\frac{\omega_\alpha\tau}{2}\ln\left(1+a^2\right),
\end{equation}
which coincides with Eq. (\ref{g_E}).

\end{appendix}


\begin{thebibliography}{10}
\bibitem{Berry84} M.V.  Berry, Proc. R. Soc. London A {\bf 392}, 45 (1984)
\bibitem{Shapere-Wilczek} A. Shapere and F. Wilczek, {\it Geometric Phases in Physics}
(World Scientific, Singapore, 1989)
\bibitem{Bohm} A. Bohm, A. Mostafazadeh, H. Koizumi, Q. Niu and J. Zwanziger, {\it The geometric
phase in quantum systems. Foundations, mathematical concepts, and
applications in molecular and condenced matter physics} (Springer, Berlin, Heidelberg, 2003)
\bibitem{Tomita} A. Tomita  and R.Y. Chiao, Phys. Rev. Lett. {\bf 57}, 937 (1986)
\bibitem{Bitter} T. Bitter and D. Dubbers, Phys. Rev. Lett. {\bf 59}, 251 (1987)
\bibitem{Wien} H. Weinfurter and G. Badurek, Phys. Rev. Lett. {\bf 64}, 1318 (1990)
\bibitem{Rauch01} Y. Hasegawa, R. Loidl, M. Baron, G. Badurek, and H. Rauch, Phys. Rev. Lett. {\bf 87}, 070401 (2001)
\bibitem{Rauch05} S. Filipp, Y. Hasegawa, R. Loidl, and H. Rauch, Phys. Rev. A {\bf 72}, 021602 (2005)
\bibitem{Rauch09} S. Filipp, J. Klepp, Y. Hasegawa, C. Plonka-Spehr, U. Schmidt, P. Geltenbort, and H. Rauch,
Phys. Rev. Lett. {\bf 102}, 030404 (2009)
\bibitem{Oberthaler} C.L. Webb, R.M. Godun, G.S. Summy, M.K. Oberthaler, P.D. Featonby, C.J. Foot, and K. Burnett,
Phys. Rev. A {\bf 60}, R1783 (1999)
\bibitem{Vedral} J.A. Jones, V. Vedral, A. Ekert,  and  G. Castagnoli,  {Nature} {\bf 403}, 869 (2000)
\bibitem{Leibfried} D. Leibfried, B. DeMarco, V. Meyer, D. Lucas, M. Barrett,
J. Britton, W.M. Itano, B. Jelenkovic, C. Langer, T. Rosenband, and
D. J. Wineland, Nature {\bf 422}, 412 (2003)
\bibitem{Aharonov59} Y. Aharonov and D. Bohm, Phys. Rev. {\bf 115}, 485 (1959)
\bibitem{Simon} B. Simon, Phys. Rev. Lett. {\bf 51}, 2167 (1983)
\bibitem{Samuel} J. Samuel  and R. Bhandari, Phys. Rev. Lett. {\bf 60}, 2339 (1988)
\bibitem{Barut} A.O. Barut, M. Bo\v{z}i\'{c}, S. Klarsfeld, and Z. Mari\'{c}, Phys. Rev. A {\bf 47}, 2581 (1993)
\bibitem{Hannay} J. H. Hannay, J. Phys. A {\bf 31}, L53 (1998)
\bibitem{Keck}  F. Keck,  H.J. Korsch,  and   S. Mossmann, J. Phys. A {\bf 36}, 2125 (2003)
\bibitem{Zwanziger} J.W. Zwanziger, S.P. Rucker, and G.C. Chingas, Phys. Rev. A {\bf 43}, 3232 (1991)
\bibitem{Moore}  D.J. Moore and G.E. Stedman, Phys. Rev. A {\bf 45}, 513 (1992)
\bibitem{Wang} S.-J. Wang, Phys. Rev. A {\bf 42}, 5107 (1990)
\bibitem{Aharonov} Y. Aharonov and J. Anandan, Phys. Rev. Lett. {\bf 58}, 1593 (1987)
\bibitem{Baudon} Ch. Miniatura, J. Robert, O. Gorceix, V. Lorent, S. Le Boiteux, J. Reinhardt, and J. Baudon,
Phys. Rev. Lett. {\bf 69}, 261 (1992)
\bibitem{Ioffe} A.G. Wagh, V.C. Rakhecha,  P. Fischer, and A. Ioffe,  Phys. Rev. Lett. {\bf 81}, 1992 (1998)
\bibitem{Sanders}  B.C. Sanders, H. de Guise, S.D. Bartlett, and W. Zhang, Phys. Rev. Lett. {\bf 86}, 369 (2001)
\bibitem{Sarandy}  M.S. Sarandy and D.A. Lidar, Phys. Rev. A {\bf 73}, 062101 (2006)
\bibitem{Zee} A. Zee, Phys. Rev. A {\bf 38}, 1 (1988)
\bibitem{Pistolesi} N. Manini and F. Pistolesi, Phys. Rev. Lett. {\bf 85}, 3067 (2000)
\bibitem{Barnett} S.M. Barnett, D. Ellinas and M.A. Dupertuis, J. Mod. Opt. {\bf 35}, 565 (1988)
\bibitem{Tewari} S.P. Tewari, Phys. Rev. A {\bf 39}, 6082 (1989)
\bibitem{Thomaz} A.C. Aguiar Pinto, M. Moutinho, and M. T. Thomaz, Braz. J. Phys. {\bf 39}, 326 (2009)
\bibitem{Chiara} G. De Chiara   and  G.M. Palma, Phys. Rev. Lett. {\bf 91}, 090404 (2003)
\bibitem{Uranyan} R.G. Unanyan and M. Fleischhauer, Phys. Rev. A {\bf 69}, 050302(R) (2004)
\bibitem{Matsumoto} X. Wang  and K. Matsumoto, J. Phys. A {\bf 34}, L631 (2001).
\bibitem{Pascazio} G. Florio, P. Facchi, R. Fazio, V. Giovannetti, and S. Pascazio, Phys. Rev. A {\bf 73}, 022327 (2006)
\bibitem{Moelmer} D. M\o{}ller, L. B. Madsen, and K. M\o{}lmer, Phys. Rev. A {\bf 75}, 062302 (2007)
\bibitem{Reich} M. Reich, U. Sterr, and W. Ertmer, Phys. Rev. A {\bf 47}, 2518 (1993)
\bibitem{Sch} W.P. Schleich, {\it Quantum Optics in Phase Space} (Wiley-VCH, Weinheim, 2001)
\bibitem{Yakovlev}  A.P. Kazantsev, G.I. Surdutovich, and V.P. Yakovlev,
{\it Mechanical Action of Light on Atoms} (World Scientific, Singapore, 1990)
\bibitem{Dalibard} J. Dalibard and C. Cohen-Tannoudji, J. Opt. Soc. Am. B {\bf 2}, 1707 (1985)
\bibitem{Oberthaler-Pfau} M.K. Oberthaler and T. Pfau, J. Phys.: Condens. Matter {\bf 15}, R233 (2003)
%\bibitem{Abramowitz} M. Abramowitz and I. A. Stegun,
%{\it Handbook of mathematical functions} (Dover Publications, Inc., New York, 1972)
\bibitem{PM} P.V. Mironova, PhD Thesis at the University of Ulm (2011)
\bibitem{Fedorov} M.V. Fedorov, {\it Atomic and free electrons in a strong light field} (World Scientific, Singapore, 1997)

\end{thebibliography}
\end{document}